\documentclass[conference,a4paper]{IEEEtran}
\IEEEoverridecommandlockouts

\IEEEpubid{%
  \makebox[\textwidth][l]{
    \parbox[t]{\columnwidth}{
      \raggedright\footnotesize
      © IFIP, 2025. This is the author’s version of the work. It is posted here by permission of IFIP for your personal use. Not for redistribution.%
    }%
  }%
}

\usepackage[a4paper,left=1.4cm,right=1.4cm,top=1.9cm,bottom=4.3cm]{geometry}

\usepackage{cite}
\usepackage{amsmath,amssymb,amsfonts}
\usepackage{algorithmic}
\usepackage{graphicx}
\usepackage{textcomp}
\usepackage{xcolor}
\usepackage[english]{babel}
\usepackage{blindtext}
\usepackage{enumitem}
\usepackage{subcaption} 
\usepackage{float} 
\usepackage[most]{tcolorbox}
\usepackage[hyphens]{url}

\def\BibTeX{{\rm B\kern-.05em{\sc i\kern-.025em b}\kern-.08em
    T\kern-.1667em\lower.7ex\hbox{E}\kern-.125emX}}
\begin{document}

\title{BBR’s Sharing Behavior with CUBIC and Reno
}

\author{
\IEEEauthorblockN{Fatih Berkay Sarpkaya, Ashutosh Srivastava, Fraida Fund, Shivendra Panwar}
\IEEEauthorblockA{\textit{NYU Tandon School of Engineering, Brooklyn, NY, USA} \\
\{fbs6417, as12738, ffund, panwar\}@nyu.edu}
}

\maketitle\IEEEpubidadjcol

\begin{abstract}

TCP BBR’s behavior has been explained by various theoretical models, and in particular those that describe how it co-exists with other types of flows. However, as new versions of the BBR protocol have emerged, it remains unclear to what extent the high-level behaviors described by these models apply to the newer versions. In this paper, we systematically evaluate the most influential steady-state and fluid models describing BBR’s coexistence with loss-based flows over shared bottleneck links. Our experiments, conducted on a new experimental platform (FABRIC), extend previous evaluations to additional network scenarios, enabling comparisons between the two models and include the recently introduced BBRv3. Our findings confirm that the steady-state model accurately captures BBRv1 behavior, especially against single loss-based flows. The fluid model successfully captures several key behaviors of BBRv1 and BBRv2 but shows limitations, in scenarios involving deep buffers, large numbers of flows, or intra-flow fairness. Importantly, we observe clear discrepancies between existing model predictions and BBRv3 behavior, suggesting the need for an updated or entirely new modeling approach for this latest version. We hope these results validate and strengthen the research community’s confidence in these models and identify scenarios where they do not apply.
\end{abstract}


\section{Introduction}

The TCP BBR congestion control algorithm is widely used in today's networks, accounting for approximately 40\% of Internet traffic by volume according to recent estimates~\cite{census}. Due to its significance, several theoretical models have been developed to better understand its behavior. With insights from these models, network operators can accurately predict traffic behavior across diverse conditions and plan protocol deployments accordingly.  However, as the protocol continues to evolve, it remains unclear to what extent the insights from existing models still apply.

One of the earliest models of BBR's behavior when competing with loss-based congestion control is the steady-state model by Ware et al.~\cite{2019IMC}. This model explains the behavior of BBRv1 in such scenarios and its accuracy was validated through experimental evaluation across a range of buffer sizes, number of flows,  link capacities and delay settings. The main findings and conclusions of this study are as follows:
\setlist{nolistsep}
\begin{enumerate}
    \item When BBRv1 flows compete with other traffic, it becomes window-limited and ACK-clocked, and sends packets at a rate determined by its inflight cap. Specifically, it aggressively pushes out loss-based competitors until it reaches its in-flight cap.
    \item The procedures for estimating the bottleneck bandwidth and the in-flight cap do not include a signal to infer the number of competing loss-based flows, nor do they adapt to achieve equal shares or fairness. Consequently, BBRv1's cap is independent of the number of competing loss-based flows.
\end{enumerate}
These findings were validated for BBRv1 against TCP CUBIC and TCP Reno in a series of experiments over the authors' own testbed, under a range of buffer settings up to 64 BDP, for selected network capacity/delay scenarios.

Several years later, Scherrer et al.~\cite{scherrer} raised a concern that investigations based on steady-state models do not generalize beyond the specific settings used in the experiment design or modeling. Fluid models allow for efficient simulation of congestion control algorithms in a wide range of scenarios. They also enable theoretical stability analysis, something not captured by steady-state models. The authors presented a fluid model for BBRv1 and BBRv2 and validated its accuracy using Mininet~\cite{mininet} experiments. Key findings include:    
\setlist{nolistsep}
\begin{enumerate}
    \item The fluid model agrees with experimental results over a range of buffer sizes from 1 to 7 BDP, with five BBR (v1 or v2) flows competing with five CUBIC or Reno flows, for selected network capacity/delay scenarios.
    \item Several key findings of previous literature are confirmed: high loss rates with BBRv1, improved loss-sensitivity with BBRv2, BBRv1's unfairness towards loss-based congestion control algorithms (CCAs), significant buffer bloat caused by BBRv1, and that BBRv2 mostly achieves its design goals of reduced buffer usage, avoiding excessive loss, and preserving fairness to loss-based CCAs.
    \item There are also novel insights regarding BBRv2 causing high buffer utilization in large drop-tail buffers.
\end{enumerate}

Although other analytical works exist~\cite{bbrdom,BBRv1-another-fluid-model-2024}, these two studies are, to the best of our knowledge, the most influential in predicting and explaining BBR’s behavior over a shared bottleneck. Their findings serve as foundational building blocks for many researchers, who often adopt the models assuming they are correct. Therefore, an independent study is essential to validate these results. Moreover, even within the same version of BBR, small heuristic changes, such as adjustments for ACK aggregation and quantization effects, could significantly alter congestion control behavior~\cite{srivastava2023some}. For example, between the 4.13 Linux kernel used by Ware et al.~\cite{2019IMC} and 5.13 used in our experiments, the BBRv1 implementation has been updated to include explicit ACK-aggregation to increase the congestion window (CWND) above the 2BDP cap in bursty ACK scenarios, the introduction of a small pacing margin below the estimated bandwidth to reduce excess queue buildup, and refining its “3 SKBs” quantization rule to ensure there is space for at least three full SKBs in flight~\cite{tcp_bbr_linux-5.13}. To identify limitations, we reevaluated these models under both their original experimental conditions and additional network settings. 

Furthermore, some authors may refer to the behavior of BBR  without specifying BBR's protocol version. It is therefore important to understand which previous conclusions, especially analytical conclusions, about these protocols apply to all versions, and which apply only to some versions. Assessing the compatibility of these models with newer BBR versions—such as applying the BBRv1 steady-state model to BBRv2 and BBRv3 or the BBRv2 fluid model to BBRv3—can provide insights into how well the core principles of earlier BBR models extend to newer versions, and highlight modifications necessary to improve existing models.

With this motivation, we evaluate and compare the performance of these models. We conducted the experiments from both papers using the FABRIC testbed \cite{baldin2019fabric}, a national-scale programmable experimental networking testbed, and also conducted experiments in additional network scenarios not considered in the original papers. Since FABRIC is an open facility, researchers can replicate and extend this study on the same testbed. We also add experimental results for BBRv3, which was introduced by Google in July 2023, one year after Scherrer et al.'s study was published and four years after Ware et al.'s study, to determine the extent to which the behavior of earlier models of BBR are still relevant for the latest protocol version.
Finally, we make all our experiment artifacts publicly available~\cite{our_artifacts_github_repo} to allow others to independently verify and build on these results.

The rest of the paper is organized as follows. In Section~\ref{background}, we provide an overview of the different BBR protocol versions and the theoretical models, including the steady-state and fluid models. Section~\ref{methodology} outlines the experimental methodology, including the topology and network settings. Section~\ref{results} presents our findings and discusses the results in detail. Finally, in Section~\ref{conclusion}, we summarize our findings and conclude the paper.

\section{Background}\label{background}

\subsection{TCP BBR protocol versions}\label{BBR models}

In this section, we summarize the main characteristics of the different BBR protocol versions, and provide some background of previous observations on BBR's behavior when sharing a bottleneck.

\textbf{BBRv1:} BBR was introduced by Google in 2016~\cite{BBRv1-2016} as a model-based congestion control algorithm that, rather than relying on packet loss, continuously estimates the network’s available bandwidth and minimum RTT. By keeping the number of in-flight packets close to the estimated bandwidth-delay product (BDP), BBR aims to achieve high throughput with a small queue. Some studies, however, found that BBRv1 could be overly aggressive against Reno or CUBIC flows, especially in shallow buffers, and since it is loss-agnostic, it experiences high packet loss rates in certain network conditions, motivating further refinements in later revisions~\cite{when-to-use-BBR,Deeper-understanding-BBR-IFIP}.

\textbf{BBRv2:} An updated version of BBR was introduced in 2019~\cite{BBRv2-IETF-104-2019} to address issues observed in BBRv1, including unfairness and high packet loss rates. Its design refines the model-based approach by adding reactions to loss or ECN signals and modifying its probing mechanisms to better share bandwidth with widely used congestion control algorithms such as CUBIC and NewReno.
However, it still has issues such as inducing higher-than-desired queueing delay when multiple BBRv2 flows coexist, and starvation of flows that experience ECN marks or early packet loss~\cite{BBRv2-problems-IETF-105}.

\textbf{BBRv3:} Because of the specific performance issues in BBRv2 such as premature exit from bandwidth probing and unfairness, BBRv3 was introduced in 2023~\cite{BBRv3-2023-July}. BBRv3 continues probing for available bandwidth until loss or ECN marking rates exceed a threshold or the bandwidth saturates. Furthermore, certain parameters of BBRv2, such as \texttt{cwnd\_gain}, \texttt{pacing\_gain}, and \texttt{inflight\_hi}, have been adjusted. However, recent studies indicate that BBRv3 still experiences challenges in bandwidth sharing—particularly among competing BBRv3 flows that start at different times or when competing against CUBIC flows in shallow-buffer scenarios, where BBRv3 tends to dominate~\cite{pam2024}.

\subsection{Theoretical Models of BBR sharing a bottleneck}\label{theoratical models}

In this study, we evaluate two theoretical models: the steady-state model by Ware~\cite{2019IMC} and the fluid model by Scherrer~\cite{scherrer}. In this section, we briefly summarize their approaches.

\textbf{Steady-State Model for BBRv1:} This model estimates BBRv1’s share of link capacity at convergence when it competes with CUBIC or Reno flows~\cite{2019IMC}. While BBRv1 is designed as a rate-based algorithm, in a congested setting it effectively becomes window-limited by its in-flight cap. By analyzing BBRv1's estimated round-trip time (RTT\textsubscript{est}) and the frequency of entering the ProbeRTT state, the model predicts BBRv1’s steady-state bandwidth share as a function of queue size, base RTT, link capacity, flow completion time after convergence, and the number of BBRv1 flows—but notably not the number of competing CUBIC or Reno flows. This explains why a single BBRv1 flow tends to get a fixed proportion of the link, regardless of how many loss-based flows are present in the bottleneck. This model predicts the BBR fraction by setting the in-flight-cap equal to the amount of data BBR flows have in-flight. In our evaluation, we test whether this in-flight cap assumption holds under various experimental conditions for BBRv1 and investigate its applicability to newer BBR versions. The formula is given below.
\begin{equation}
    p = \frac{1}{2} - \frac{1}{2X} - \frac{4N}{q}
\end{equation}

\begin{equation}
    BBR_{frac} = (1 - p) \times \left( \frac{d - \text{Probe}_{time}}{d} \right)
\end{equation}

where \( BBR_{frac} \) is the link capacity share of BBR flows, \( p \) is the aggregate fraction of the loss-based flows (i.e., CUBIC or Reno), \( X \) is the buffer size in BDP, \( N \) is the number of BBR flows in the bottleneck, \( q \) is the buffer size in packets, \( d \) is the duration after convergence, and \(\text{Probe}_{time}\) is the time spent in the ProbeRTT state~\cite{2019IMC}.

\textbf{Fluid Model for BBRv1 and BBRv2:} Scherrer et al.~\cite{scherrer} develop a fluid-model framework that captures the core mechanisms of BBRv1 and BBRv2 using differential equations. Their model explicitly tracks the estimated bottleneck bandwidth (\text{BtlBw}) and minimum RTT (\text{RTprop}), along with the transitions between BBR’s key phases, ProbeBW and ProbeRTT. They approximate BBR’s probing pulses with smooth sigmoidal functions and use discrete mode variables to represent the state machine, effectively transforming BBR’s stepwise behavior into continuous dynamics. Simulating these equations enables the prediction of queue lengths, flow rates, packet losses, and jitter. In the model, the sending rate is formulated as follows, accounting for the switching between the ProbeRTT and ProbeBW phases:

\begin{equation}
x_i \;=\; m_i^{\mathrm{prt}} \cdot \frac{w_i^{\mathrm{prt}}}{\tau_i}
\;-\;\bigl(1 - m_i^{\mathrm{prt}}\bigr)\,\cdot\,x_i^{\mathrm{pbw}}
\end{equation}

This equation represents the sending rate \( x_i \) of a BBR flow, which depends on whether the flow is in the \textit{ProbeRTT} or \textit{ProbeBW} phase. The binary variable \( m_i^{\mathrm{prt}} \) determines the mode: when \( m_i^{\mathrm{prt}} = 1 \), the flow is in \textit{ProbeRTT}, and its rate is constrained by the inflight limit \( w_i^{\mathrm{prt}} \) divided by the round-trip time \( \tau_i \), effectively reducing the sending rate to minimize queueing delay. Conversely, when \( m_i^{\mathrm{prt}} = 0 \), the flow operates in \textit{ProbeBW}, where the sending rate follows \( x_i^{\mathrm{pbw}} \), which is dictated by BBR's bandwidth estimation and pacing mechanisms. The equation transitions between these two phases to capture BBR's state-dependent behavior. Through our experiments, we assess how accurately this approach captures the behavior of BBRv2 and investigate its applicability to BBRv3.

\section{Experiment Methodology}\label{methodology}

We systematically evaluate the models through controlled experiments. This section details our methodology.

\textbf{Experiment Platform:} We conduct all our experiments on FABRIC \cite{baldin2019fabric}, a national scale programmable experimental networking testbed. Because it is a shared facility, using FABRIC allows others to more easily reproduce our work and reuse our experiment artifacts. 

\begin{figure}[htbp]
    \centering
    \includegraphics[width=0.35\textwidth]{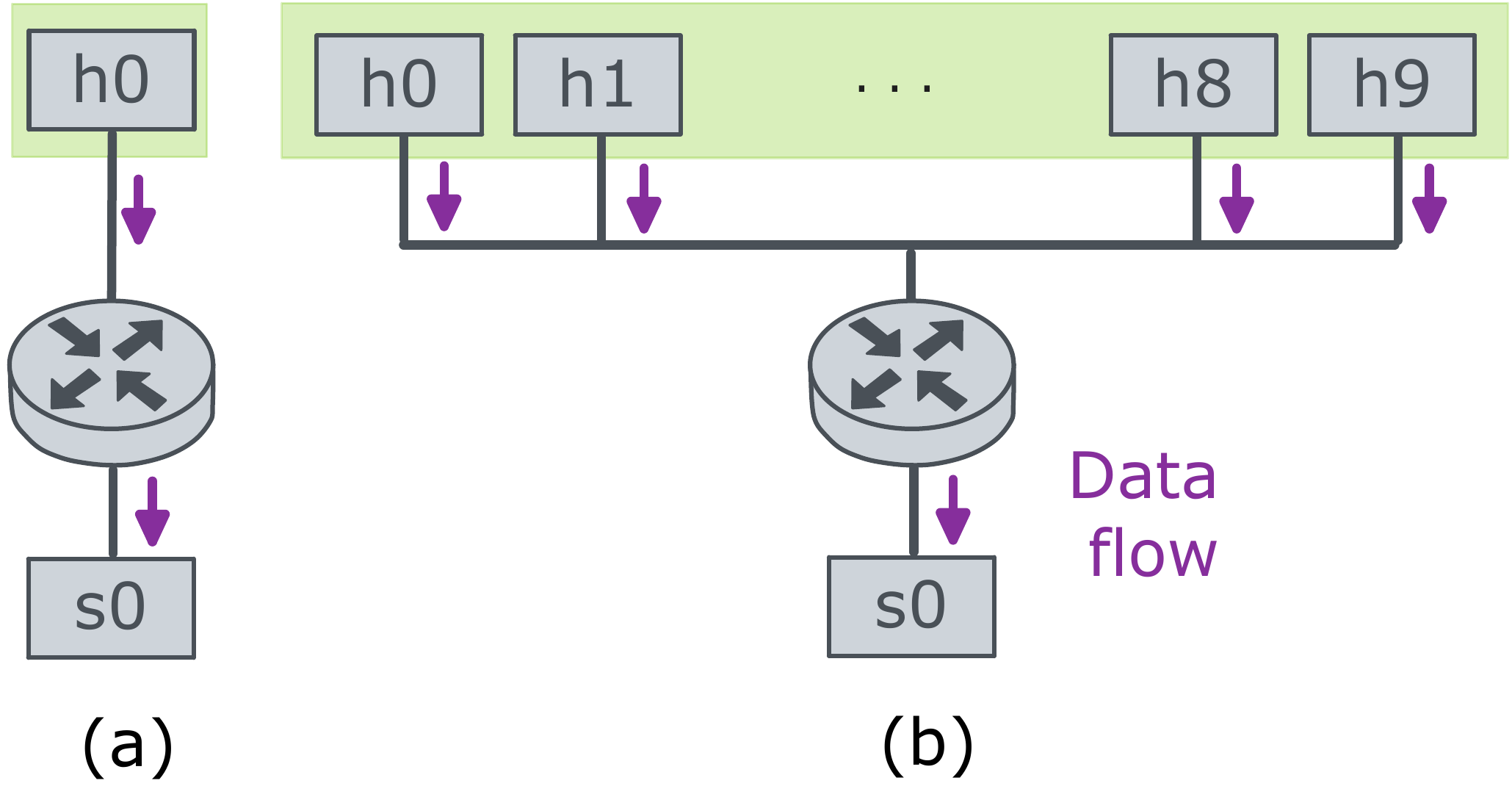}
    \caption{Topologies.}
    \label{fig: topology}
\end{figure}

\textbf{Topology:} We have two different experiment sets with two different topologies. In one of them, as seen in Fig.~\ref{fig: topology}a, there is a single sending host and a single receiving host on either side of the router, similar to Ware et al.~\cite{2019IMC}. In another one, as shown in Fig.~\ref{fig: topology}b, there are ten sending hosts and a single receiving host. Each node in our experiment is a virtual machine running Ubuntu 22.04, with 4 cores and 32 GB RAM. 

\textbf{Network Settings:} Following the original papers, we have different network scenarios for the two studies.

From Ware et al.~\cite{2019IMC}, we have the following  scenarios (refer to Fig.~\ref{fig: topology}a.):
\begin{itemize}
\item 40 ms base RTT, 10 Mbps bottleneck link capacity 
\item 30 ms base RTT, 50 Mbps bottleneck link capacity 
\end{itemize}
with bottleneck buffer sizes in multiples of the link bandwidth delay product (BDP): 1, 2, 4, 8, 16, 32, 64. 

In deep buffer size settings (16, 32, and 64 BDP), we had to increase the socket buffer sizes from 16MB (default) to 64MB at the end hosts. This adjustment was necessary; otherwise, we observed that the flows became application-limited or receiver-window-limited most of the time during the connection.

Following Scherrer et al.~\cite{scherrer} (refer to Fig.~\ref{fig: topology}b.), we use a 100 Mbps bottleneck capacity with a fixed 10 ms delay on the receiver side and a randomized sender delay to achieve a total RTT uniformly distributed between 30 and 40 ms. In the original paper, the bottleneck buffer sizes are from 1 to 7 BDP. However, as mentioned in Ware et al.~\cite{2019IMC}, \cite{ware_reference_high_buffer} suggests that home routers can have deep buffers. Therefore, our analysis should not be limited to only shallow buffers. Accordingly, we also consider some buffer sizes from 8 to 64 BDP, extending the original findings of the Scherrer et al. paper.

To configure the bottleneck bandwidth and buffer size, we use the token bucket filter implemented in \texttt{tc-htb} at the egress interface of the router. We use \texttt{netem} to emulate delay.

We disable network interface card (NIC) offloading features, and we use the default MTU of 1500 bytes on all interfaces. The purpose is to mimic the behavior of flows traversing consumer-grade routers.

\textbf{Congestion Control Implementations:} We conduct experiments with BBRv1, CUBIC, and Reno, using their implementation in Linux kernel 5.13.12. For BBRv2, we use the \texttt{v2alpha} branch of the official BBR repository~\cite{bbrrepo}, and for BBRv3 we use the \texttt{v3} branch in Linux kernel 6.4.0.

\textbf{Flow Generation}: We generate flows from sender to receiver using the \texttt{iperf3} utility, with flow durations as follows:

\begin{itemize}
    \item In Ware et al.~\cite{2019IMC}, BBR vs. CUBIC coexistence was analyzed for 400 seconds, and BBR vs. Reno for 200 seconds post-convergence. We run flows for 1000 seconds in total, analyzing the final 400 or 200 seconds for CUBIC and Reno, respectively. (refer to Fig.~\ref{fig: topology}a.)
    \item The evaluation in Scherrer et al.~\cite{scherrer} uses a flow duration of 9 seconds, and reports results for the last 5 seconds. In our FABRIC testbed experiments, we observed that this duration is too small to observe flow convergence. Hence, we chose the flow duration to be 5 minutes, and report results for the last 2 minutes. (refer to Fig.~\ref{fig: topology}b.) 
\end{itemize}

We use bulk TCP flows to remain consistent with the methodologies of the original studies~\cite{2019IMC, scherrer}, allowing direct comparison with prior work.

\textbf{Trials}: Following~\cite{scherrer}, each experiment is repeated three times; we report the mean and standard error across trials.

\textbf{Models:} We compare experimental results to the predictions of the steady state model described by Ware et al.~\cite{2019IMC} and the fluid model of Scherrer et al.~\cite{scherrer}. For the steady state model, we wrote our own implementation using the formulations in the original paper, and for the fluid model we used the implementation in their shared artifacts~\cite{2022_github_repo}. 

For the settings from Ware et al.\cite{2019IMC}, we cannot precisely replicate long-running, post-convergence scenarios with the fluid model due to excessive RAM usage for many flows or long durations. Instead, we run the fluid model for $9$ seconds and analyze the final $5$ seconds, as in Scherrer et al.\cite{scherrer}.

\begin{figure*}[t] 
  \centering
  \begin{subfigure}[b]{0.31\textwidth} 
      \centering
      \includegraphics[width=\textwidth]{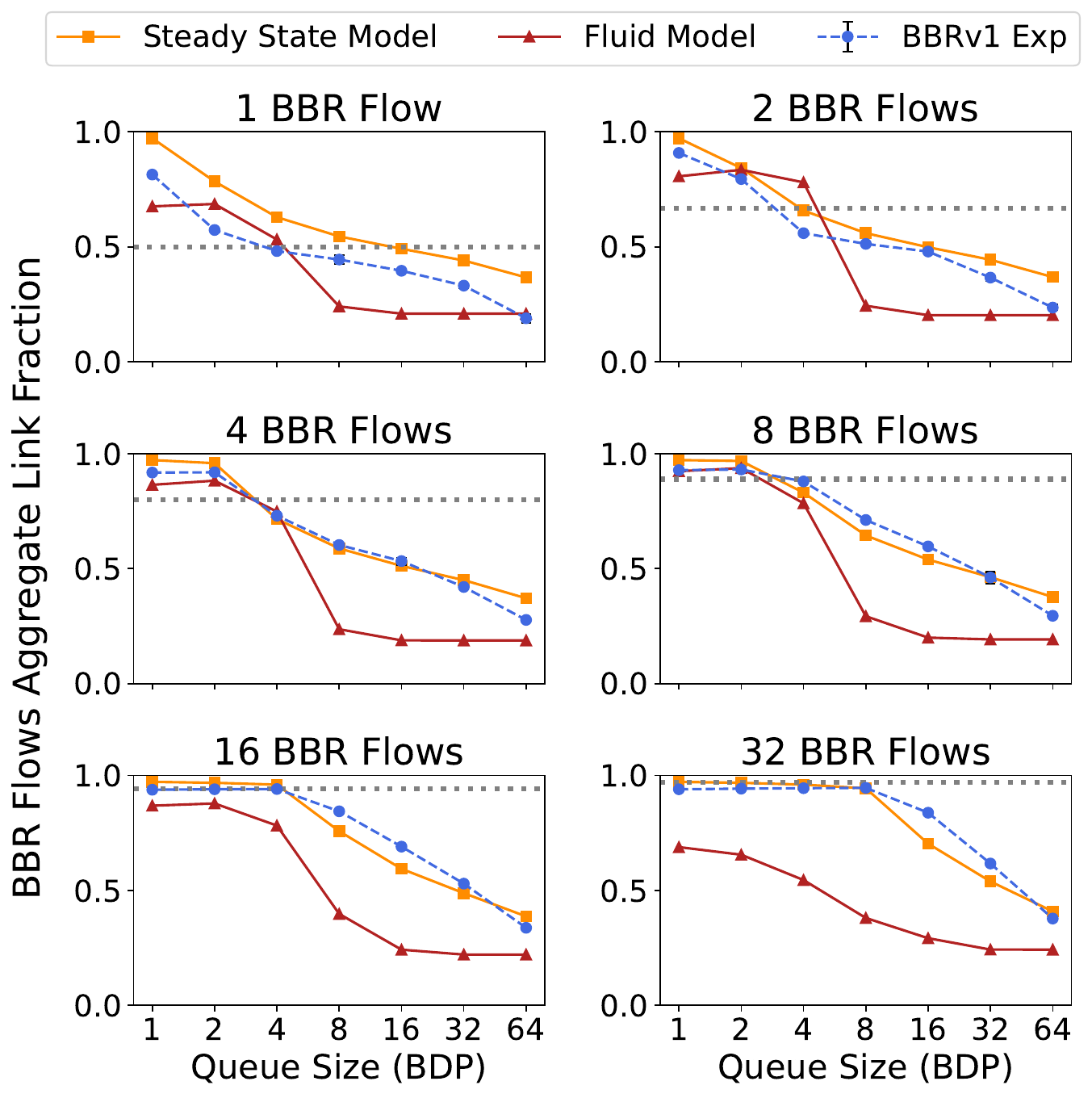}
      \caption{40ms × 10 Mbps, BBRv1 vs 1 CUBIC (MSE is 0.0067 for the steady state model,  0.0650 for the fluid model.)}
      \label{fig:part1-bbrv1-a}
  \end{subfigure}
  \hfill 
  \begin{subfigure}[b]{0.31\textwidth}
      \centering
      \includegraphics[width=\textwidth]{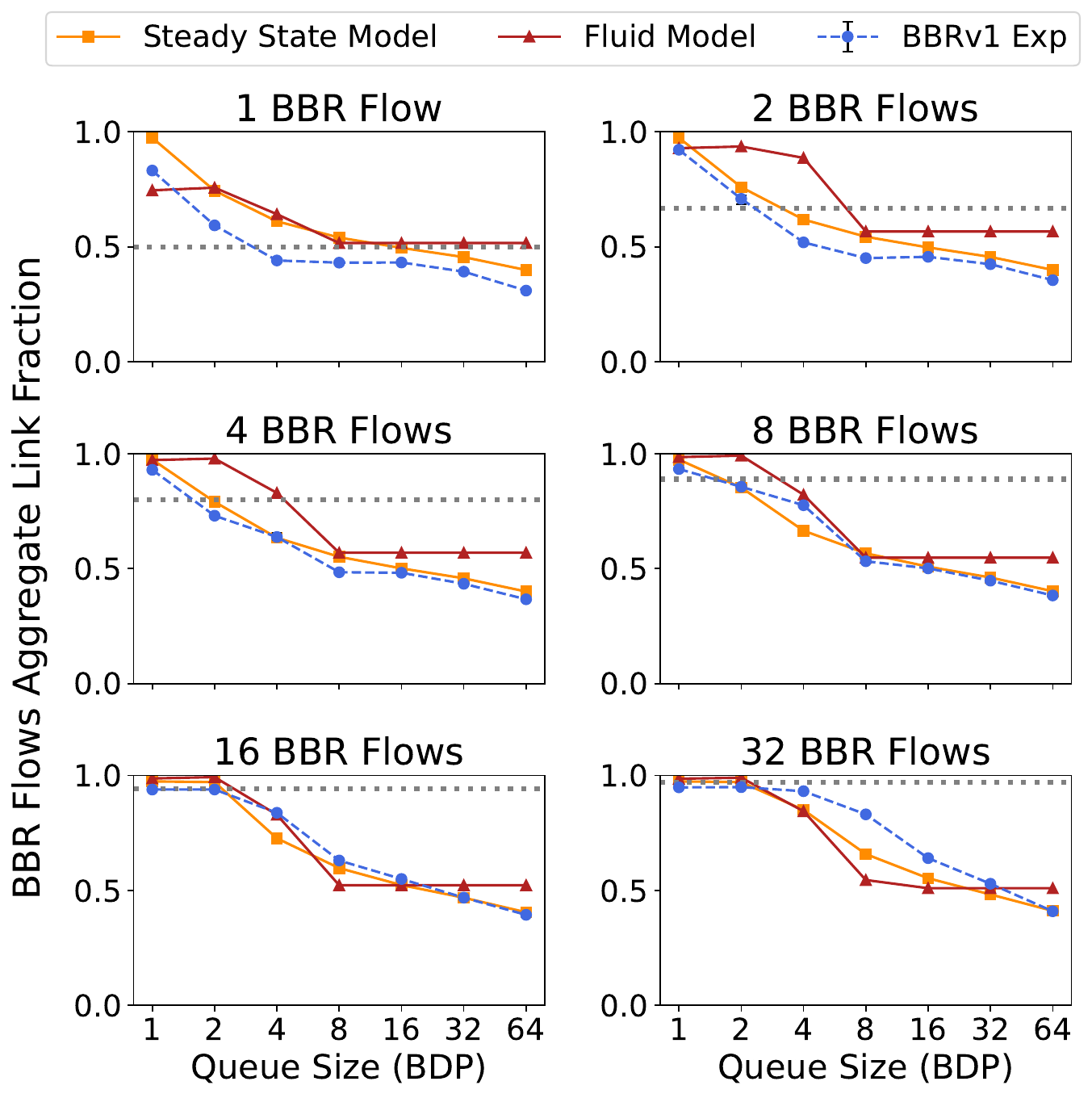}
      \caption{30ms × 50 Mbps, BBRv1 vs 1 CUBIC (MSE is 0.0052 for the steady state model,  0.0195 for the fluid model.)} 
      \label{fig:part1-bbrv1-b}
  \end{subfigure}
  \hfill 
  \begin{subfigure}[b]{0.31\textwidth}
      \centering
      \includegraphics[width=\textwidth]{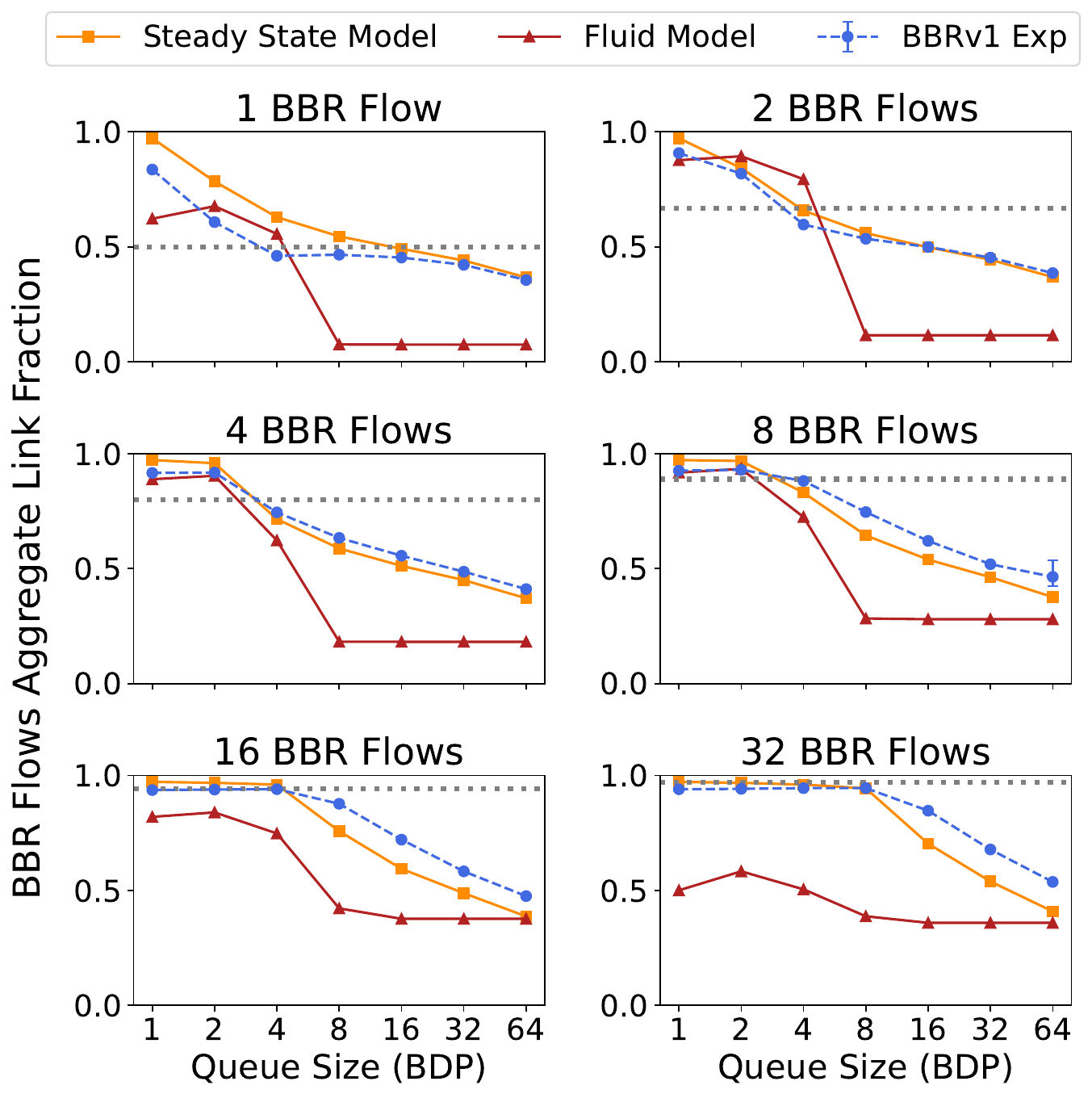}
      \caption{40ms × 10 Mbps, BBRv1 vs 1 Reno (MSE is 0.0060 for the steady state model,  0.0880 for the fluid model.)}
      \label{fig:part1-bbrv1-c}
  \end{subfigure}
  \caption{Aggregate link fraction captured by BBRv1 flows sharing a bottleneck with a loss-based flow. The gray dashed line indicates fair sharing. Experimental results are compared with predictions from the steady-state model~\cite{2019IMC} and fluid model~\cite{scherrer}.} 
  \label{fig:part1-bbrv1}
\end{figure*}

\begin{figure*}[t] 
  \centering
  \begin{subfigure}[b]{0.31\textwidth} 
      \centering
      \includegraphics[width=\textwidth]{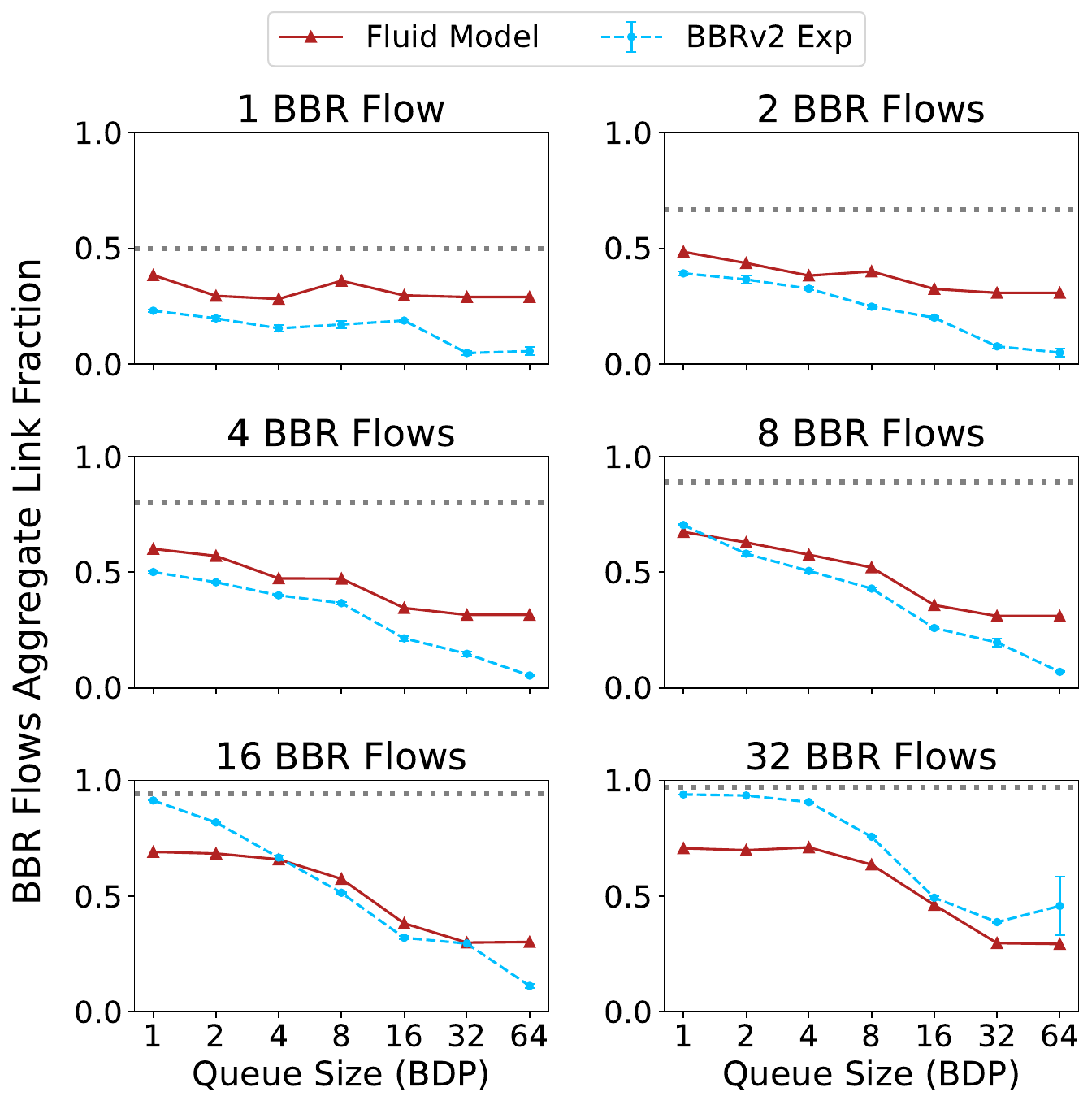}
      \caption{40ms × 10 Mbps, BBRv2 vs 1 CUBIC (MSE is 0.0225)}
      \label{fig:part1-bbrv2-a}
  \end{subfigure}
  \hfill 
  \begin{subfigure}[b]{0.31\textwidth}
      \centering
      \includegraphics[width=\textwidth]{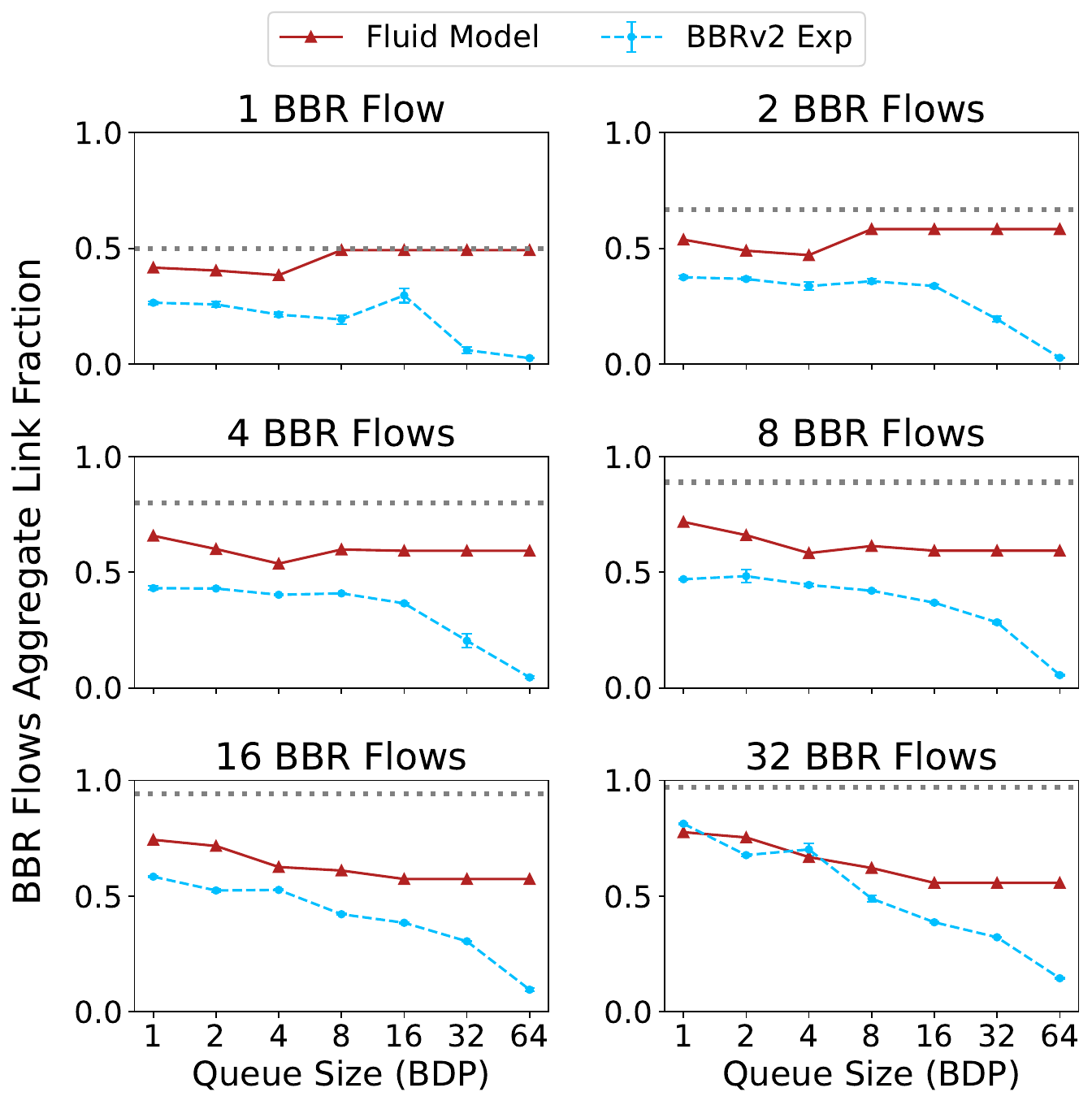}
      \caption{30ms × 50 Mbps, BBRv2 vs 1 CUBIC (MSE is 0.0757)}
      \label{fig:part1-bbrv2-b}
  \end{subfigure}
  \hfill 
  \begin{subfigure}[b]{0.31\textwidth}
      \centering
      \includegraphics[width=\textwidth]{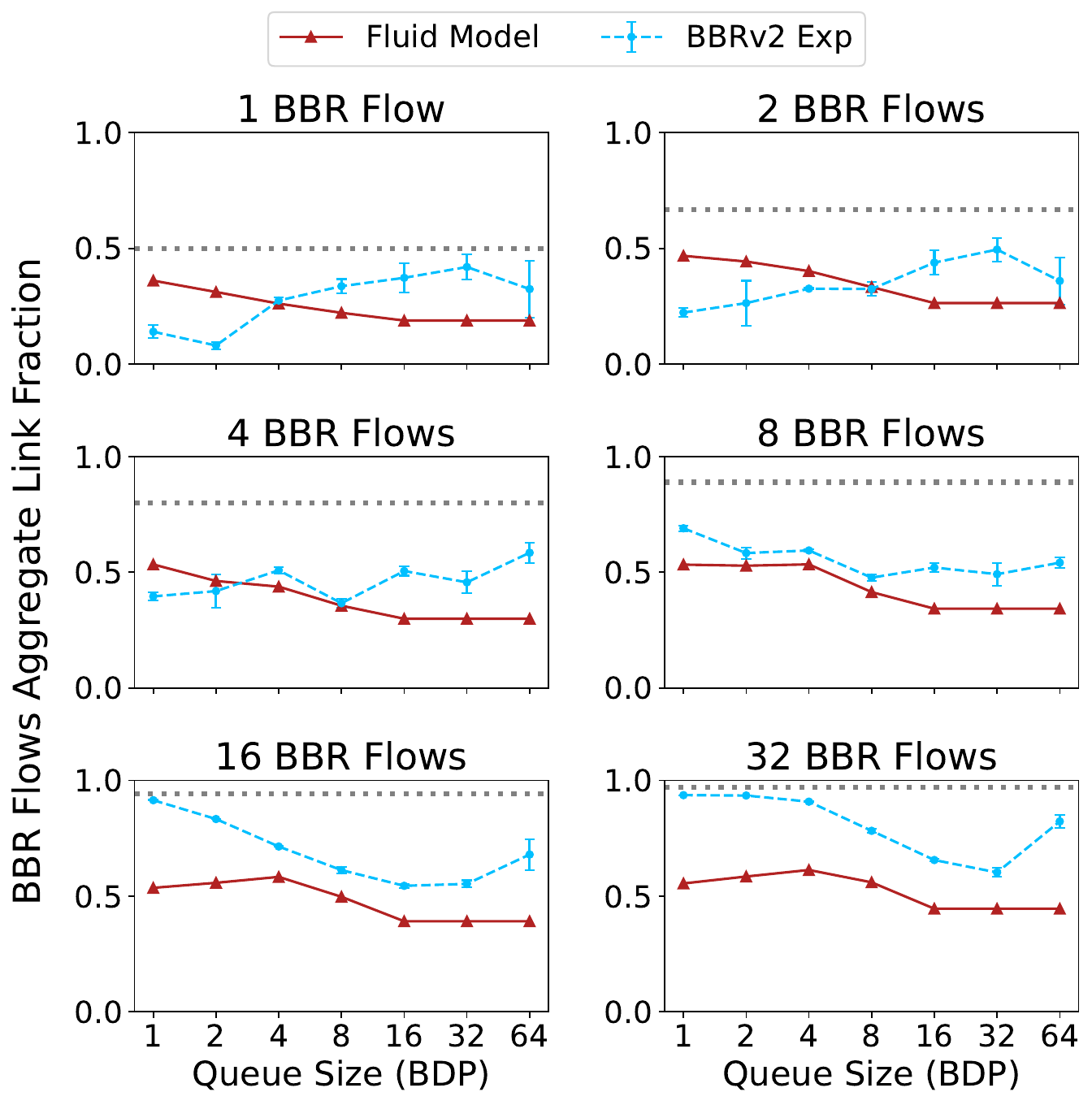}
      \caption{40ms × 10 Mbps, BBRv2 vs 1 Reno (MSE is 0.0408)}
      \label{fig:part1-bbrv2-c}
  \end{subfigure}
  \caption{Aggregate link fraction captured by BBRv2 flows sharing a bottleneck with a loss-based flow. The gray dashed line indicates fair sharing. We compare the predictions of the fluid model for BBRv2~\cite{scherrer} to experimental results.} 
  \label{fig:part1-bbrv2}
\end{figure*}

\begin{figure*}[!htbp]
    \centering
    \includegraphics[width=0.99\textwidth]{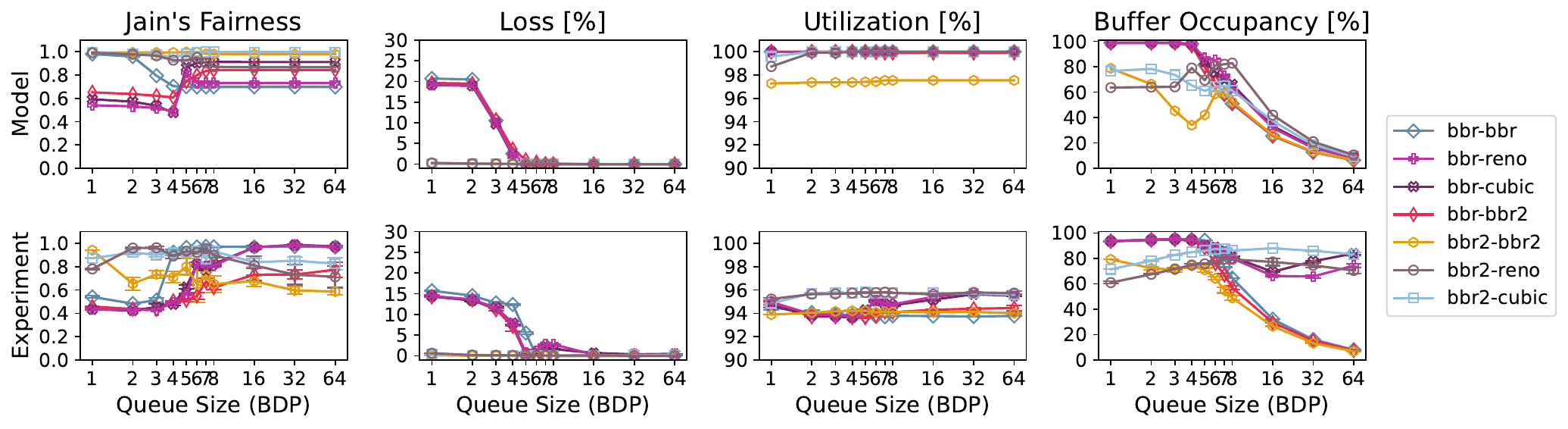}
    \caption{Fluid model prediction and experiment results when BBR shares a bottleneck with an equal number of flows of another type (different BBR version or loss-based CC), or when all flows sharing the bottleneck are the same.}
    \label{fig:part1-2022-exps}
\end{figure*}

\section{Experimental Results}\label{results}

In this section, we present the results of our experiments and compare our findings with those from the original papers, as well as with their corresponding theoretical models.

We organize this section into three parts.

First, we evaluate BBRv1 and BBRv2 against single loss-based flows, CUBIC and Reno, using the topology shown in Fig.~\ref{fig: topology}a. For BBRv1, we validate Ware et al.’s findings~\cite{2019IMC} and compare its steady-state model with Scherrer et al.’s fluid model~\cite{scherrer} to determine the conditions under which each model can predict BBR behavior. Similarly, we evaluate BBRv2 using Scherrer et al.’s fluid model.

Second, using the topology in Fig.~\ref{fig: topology}b, we compare our experimental results for BBRv1 and BBRv2 against their respective fluid models when competing with multiple loss-based or BBR flows.

Third, by using the same topologies, we evaluate BBRv3 against single and multiple flows, highlighting its improvements and weaknesses compared to previous BBR versions. While neither the steady-state model~\cite{2019IMC} nor the fluid model~\cite{scherrer} was designed for BBRv3, BBRv3 resembles BBRv1 and BBRv2. Thus, we investigate whether these models can capture its behavior to any extent by comparing their predictions with experimental results.

\subsection{Evaluating BBRv1 \& BBRv2 against single loss-based flow}\label{sec:validating results}

\textbf{BBRv1 vs a single loss-based flow:} The key findings from Ware et al.~\cite{2019IMC} relate to BBRv1's aggressiveness when competing with loss-based CCAs. We confirm these findings in Fig.~\ref{fig:part1-bbrv1}, which shows the aggregate fraction of the link capacity obtained by BBRv1 flows when competing with a loss-based flow, for a variety of link settings and bottleneck buffer sizes. BBRv1 captures more than its fair share of the link capacity for shallow buffers, and less than its share for deep buffers. As in Ware et al.~\cite{2019IMC}, we have good agreement between the experimental results and the steady state model proposed. This agreement confirms the core motivation behind the model and validates its key assumption: ``BBRv1’s sending rate is limited by its in-flight cap when competing with loss-based flows." The throughput share of BBRv1 with loss-based CCAs is well-documented in the literature and our results confirm the previous findings in this setting~\cite{understanding-of-bbrv2-bbrv1,2019IMC,scherrer}.

For comparison, we also apply the fluid model of Scherrer et al.~\cite{scherrer} to the same scenarios. For a small number of flows, and shallow to moderately deep buffers, the fluid model also has good agreement with the experiment results. However, in scenarios with deep buffers and a large number of BBR flows, we observe significant discrepancies. We suspect that these disagreements may be due to the limited duration of the fluid model simulation, as 9 seconds could be too short to reach convergence in such cases.\footnote{Due to excessive RAM usage with long durations, we limit the fluid model to 9 seconds and analyze the final 5 seconds as in the original study~\cite{scherrer}.} As shown in Ware et al.\cite[Figure 2]{2019IMC}, convergence time is longer in deep buffers compared to shallow ones. Additionally, with a high number of flows, convergence time increases because BBR flows take longer to stabilize as they compete for bandwidth and adapt to network conditions.

In the original paper~\cite{scherrer}, the authors only evaluated the scenario of 5 BBRv1 flows competing with 5 CUBIC/ RENO flows, and only considered buffer sizes up to 7 BDP. With this evaluation, we show the weakness of the fluid model in scenarios with large BDP sizes and for a large number of competing flows. 

Overall, to provide a more quantitative assessment of the model’s predictive accuracy, the average mean squared error (MSE) of the steady-state model is 0.006 (average root mean squared error, RMSE: 0.07), while that of the fluid model is 0.058 (average RMSE: 0.22).

\begin{tcolorbox}[beforeafter skip=0.5\baselineskip, before upper={\parindent15pt},colback=black!3!white,colframe=black,
]

The Ware model~\cite{2019IMC} accurately predicts BBRv1 behavior against a single loss-based flow across a wide range of buffer settings. The fluid model~\cite{scherrer} is less effective with a large number of flows and deep buffers.

\end{tcolorbox}

\textbf{BBRv2 vs a single loss-based flow:} Unlike BBRv1, BBRv2 captures less than its fair share of the link capacity in both shallow and deep buffers as shown in Fig.\ref{fig:part1-bbrv2}. BBRv2 is loss-sensitive and includes the additional \texttt{in\_flight\_hi} mechanism, making it is less aggressive~\cite{understanding-of-bbrv2-bbrv1,pam2024,scherrer}. Since loss-based algorithms are buffer-filling, BBRv2 experiences excessive packet losses, which cause its \texttt{in\_flight\_hi} to get stuck at lower values, preventing it from increasing its sending rate~\cite{bbrv1-vs-bbrv2-examining-performance-differences}. When competing against CUBIC, the loss-based flows dominates BBRv2 in deep buffers. However, BBRv2 achieves a larger share of the bottleneck bandwidth in deep buffers when competing against Reno (Fig.~\ref{fig:part1-bbrv2-c}), likely due to Reno's lower aggressiveness. To the best of our knowledge, this difference in BBRv2's coexistence with CUBIC and Reno is not documented in the literature.

The fluid model successfully captures the reduced aggressiveness of BBRv2 in Fig.~\ref{fig:part1-bbrv2}, compared to BBRv1 in Fig.~\ref{fig:part1-bbrv1}. It accurately predicts that BBRv2 captures less than its fair share of the bottleneck capacity. However, it fails to reflect key details from our experimental results. First, the model overestimates BBRv2’s link share when competing against CUBIC. Second, it does not effectively distinguish between Reno’s and CUBIC’s aggressiveness, underestimating BBRv2's bandwidth advantage over Reno, especially in deep buffers and with many flows. This may be due to the fluid model's short simulation duration, which might be insufficient for full convergence. 

More quantitatively, the average MSE of the fluid model is 0.046 (average RMSE: 0.21). Although not shown here, we also evaluated Ware’s steady-state model~\cite{2019IMC}, which was not designed for this protocol and, it performs poorly, with an average MSE of 0.1 (average RMSE: 0.31).

\begin{tcolorbox}[beforeafter skip=0.5\baselineskip, before upper={\parindent15pt},colback=black!3!white,colframe=black,
]

BBRv2 is less aggressive than BBRv1 and gets less than its fair share of capacity against loss-based flows, in line with the fluid model predictions~\cite{scherrer}. However, the fluid model overestimates BBRv2's link share when competing against CUBIC and underestimates it when competing against Reno, especially in scenarios involving deep buffers and numerous flows.

\end{tcolorbox}

\subsection{Evaluating BBRv1 and BBRv2 against multiple flows}\label{BBR_multiple_flows}

The model of Ware et al.~\cite{2019IMC} is specific to the setting of a single loss-based flow competing with BBRv1 flows. More recent results by Scherrer et al.\cite{scherrer} also examine the intra-CCA fairness of both BBRv1 and BBRv2. In this section, we consider the network setting of Scherrer et al., where either:
\begin{itemize}
    \item an equal number of flows of two different types (BBR vs. loss based, or different BBR protocol versions),
    \item or only BBR flows of the same protocol version,
\end{itemize} 
compete over a shared bottleneck.

Fig.~\ref{fig:part1-2022-exps} shows results for Jain's fairness index (JFI), loss rate, link utilization, and buffer occupancy when BBR flows compete with other BBR or loss-based flows over a shared bottleneck with a range of buffer sizes. 

In this figure, the first row reproduces the fluid model using publicly available artifacts from Scherrer et al.~\cite{scherrer}, while the second row shows our experimental results for their settings, and includes additional results for deep buffers.

Our experiment results mostly agree with the fluid model and experiment results of Scherrer et al., especially regarding loss for all cases, and fairness between BBR and loss-based flows. However we note several differences for other cases.

\textbf{Intra-flow Fairness:} (Refer to the first column in Figure~\ref{fig:part1-2022-exps}.) In the BBRv1 vs BBRv1 case, we observe poor fairness at 1-3 BDPs in our experiments, but the fairness improves as the buffer size increases to 7 BDP. This contradicts both the fluid model predictions (good fairness at 1 BDP, and the fairness becomes worse as the buffer size increases) and the original study's experimental results (good fairness for all buffer sizes from 1 to 7 BDP). 

The discrepancy with the fluid model is already explained in the original study~\cite{scherrer}. In short, the model exaggerates RTT unfairness, whereas in practice, random noise compensates for the difference, leading to better fairness than predicted. For unfairness in shallow buffers, a possible explanation is BBRv1’s aggressiveness. Despite losses, flows continue sending aggressively, and with a high number of flows, some experience more drops than others, leading to unequal bandwidth sharing. In deep buffers, this effect diminishes because all flows’ sending rates are limited by their in-flight cap, and losses are minimal.

In the BBRv2 vs. BBRv2 case, fairness is better in shallow buffers because BBRv2 flows set their \texttt{in\_flight\_hi} cap due to packet loss which limits the in-flight data. However, in deep buffers, our results show poor fairness, contradicting both the model predictions and the original studies' experimental results. A possible cause is the de-synchronization of BBR flows and their different RTTs. In our experiment, the flows may not enter the bottleneck at the same time, Then, the earlier flows could obtain a higher \texttt{in\_flight\_hi} and keeps increasing its sending rate, while later flows remain constrained due to their lower \texttt{in\_flight\_hi}, which prevents them from increasing their delivery rate. 

Previous studies have also observed this difference in intra-fairness performance between BBRv1 and BBRv2 in deep and shallow buffers~\cite{bbrv1-vs-bbrv2-examining-performance-differences,exp-evaluation-BBRv1,intra-protocol-problem-BBRv2,understanding-of-bbrv2-bbrv1}.

\textbf{Inter-flow Fairness:} When BBRv1 competes against loss-based flows (or against BBRv2), fairness is low in shallow buffers but improves as the buffer size increases. In contrast, BBRv2 generally shows better fairness with loss-based flows. 
These findings are consistent with our previous experiments, as shown in Figures~\ref{fig:part1-bbrv1} and \ref{fig:part1-bbrv2}.

\textbf{Loss:} (Refer to the second column in Fig.~\ref{fig:part1-2022-exps}.) We confirm that BBRv1 causes high loss in shallow buffers, consistent with the fluid model and the original experimental results\cite{scherrer}.

\textbf{Utilization:} (Refer to the third column in Fig.~\ref{fig:part1-2022-exps}.) For buffer sizes up to 7 BDP, our results differ from both the original experiment and the model in absolute value. This discrepancy may be due to differences in utilization measurement or link capacity settings (e.g., which protocol overheads are included). We observe lower utilization when only BBR flows are present compared to scenarios with a mix of BBR and loss-based flows. This is expected, as loss-based TCP continuously fills the buffer, while BBR flows have synchronized ProbeRTT phases causing temporary underutilization. 

\textbf{Buffer Occupancy:} (Refer to the fourth column in Fig.~\ref{fig:part1-2022-exps}.) For buffer sizes up to 7~BDP, we confirm the original finding that BBRv1 tends to fill the bottleneck buffer, even though its design aims to minimize buffer occupancy. This occurs both when all flows are BBRv1 and when BBRv1 shares the bottleneck with other flows.

When all competing flows are BBRv2, ~\cite{scherrer} found that buffer occupancy decreases first and then increases as buffer size goes from 1 BDP to 7 BDP. In contrast, our results show a gradual decrease in buffer occupancy from 1 to 7 BDP.

For large buffer sizes, the fluid model predicts low buffer occupancy when BBR (any version) competes against loss-based congestion control. However, our experiments indicate high buffer occupancy in these cases. In our experiments, BBR achieves low delay benefits only when it does \textbf{not} share a buffer with a loss-based congestion control, consistent with other published results (e.g.,~\cite{bbrdom}). This discrepancy further highlights the fluid model’s limitations.

\begin{tcolorbox}[beforeafter skip=0.5\baselineskip, before upper={\parindent15pt},colback=black!3!white,colframe=black,
]
 
While the fluid model~\cite{scherrer} aligns with our results for loss and inter-flow fairness for BBRv1 and BBRv2, it fails to capture critical behaviors  - particularly intra-flow fairness, buffer occupancy, and utilization - when buffer sizes are very large. 

\end{tcolorbox}

\begin{figure*}[t] 
  \centering
  \begin{subfigure}[b]{0.31\textwidth} 
      \centering
      \includegraphics[width=\textwidth]{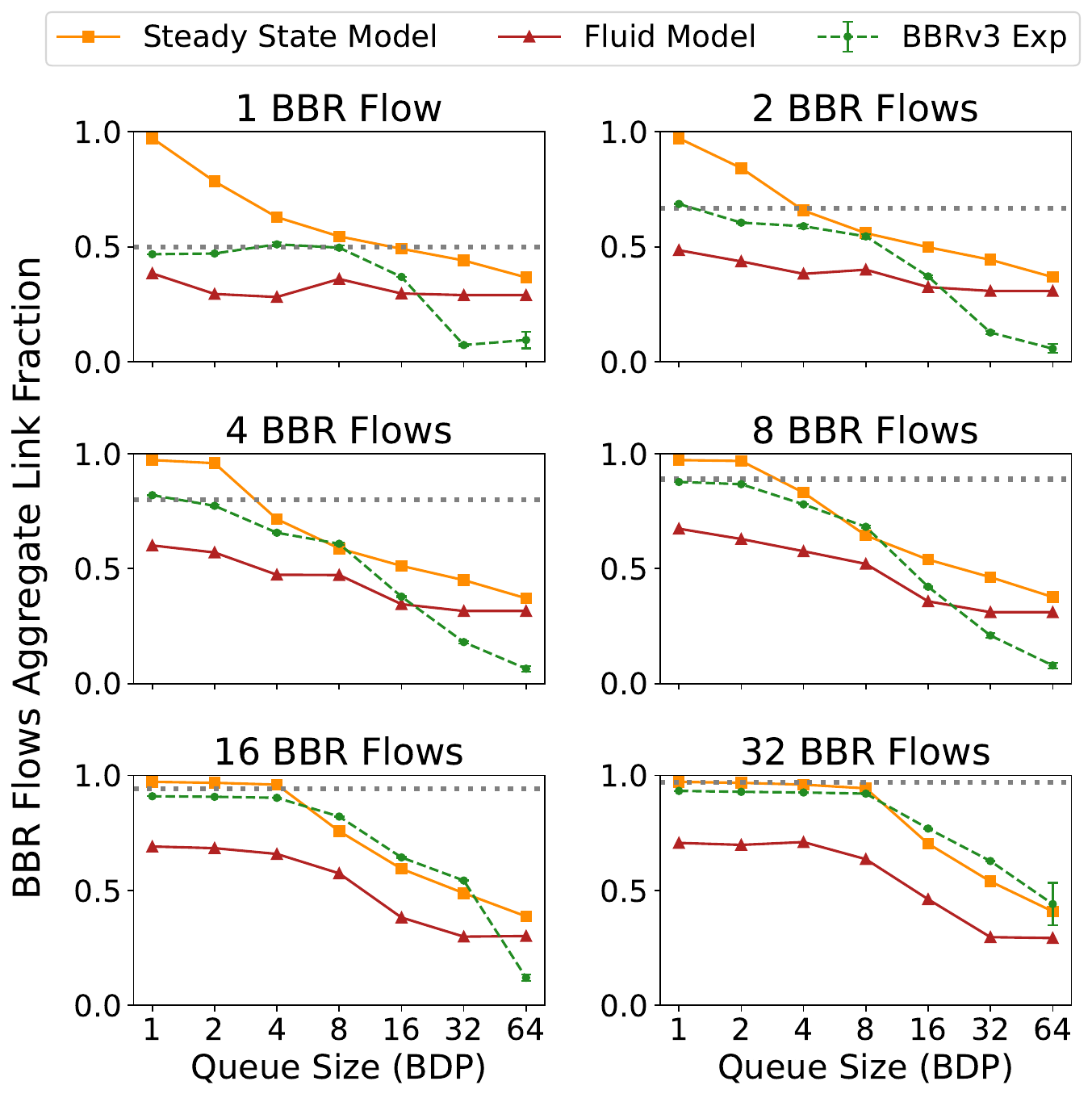}
      \caption{40ms × 10 Mbps, BBRv3 vs 1 CUBIC}
      \label{fig:part2-bbrv3-a}
  \end{subfigure}
  \hfill 
  \begin{subfigure}[b]{0.31\textwidth}
      \centering
      \includegraphics[width=\textwidth]{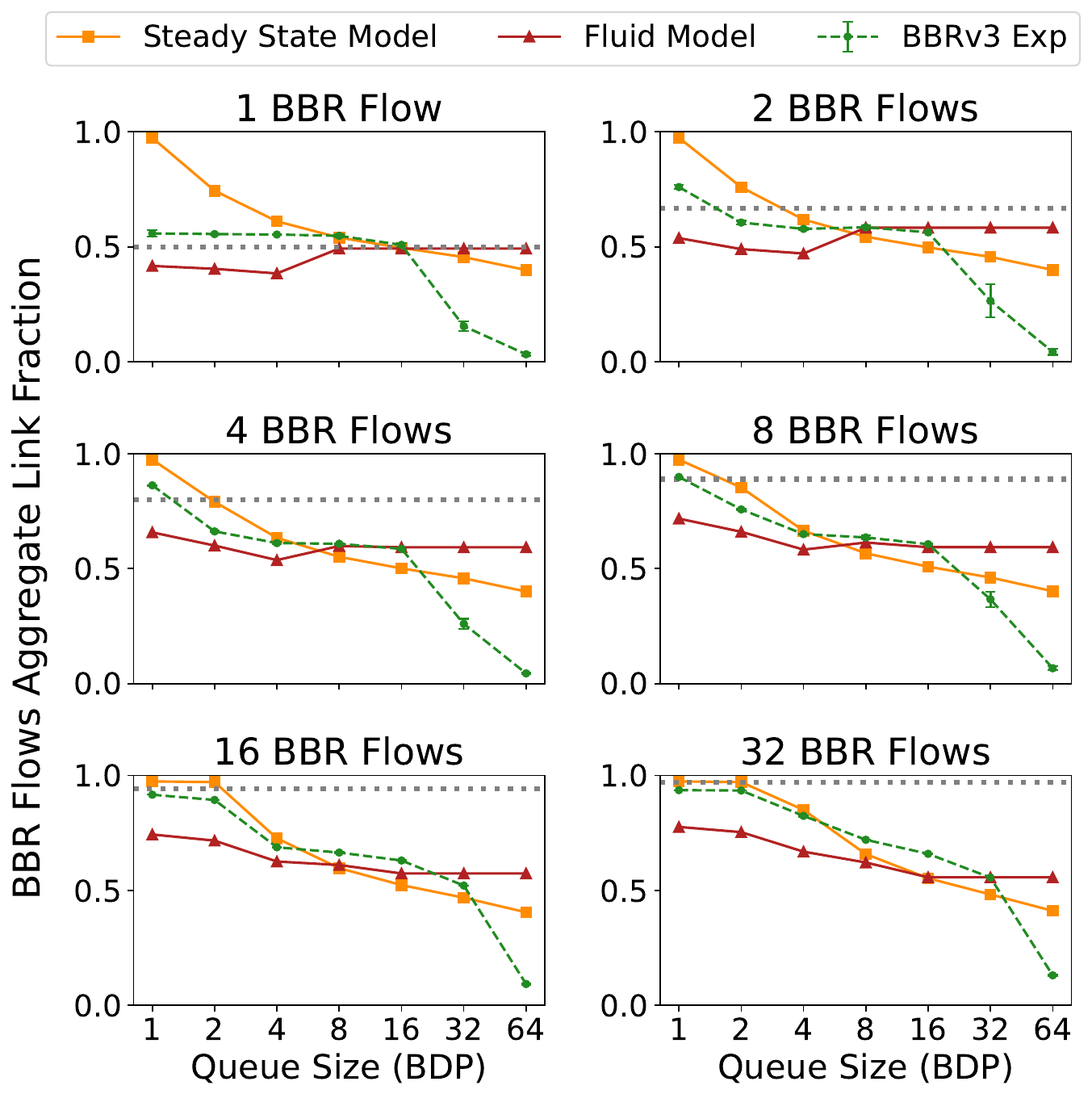}
      \caption{30ms × 50 Mbps, BBRv3 vs 1 CUBIC}
      \label{fig:part2-bbrv3-b}
  \end{subfigure}
  \hfill 
  \begin{subfigure}[b]{0.31\textwidth}
      \centering
      \includegraphics[width=\textwidth]{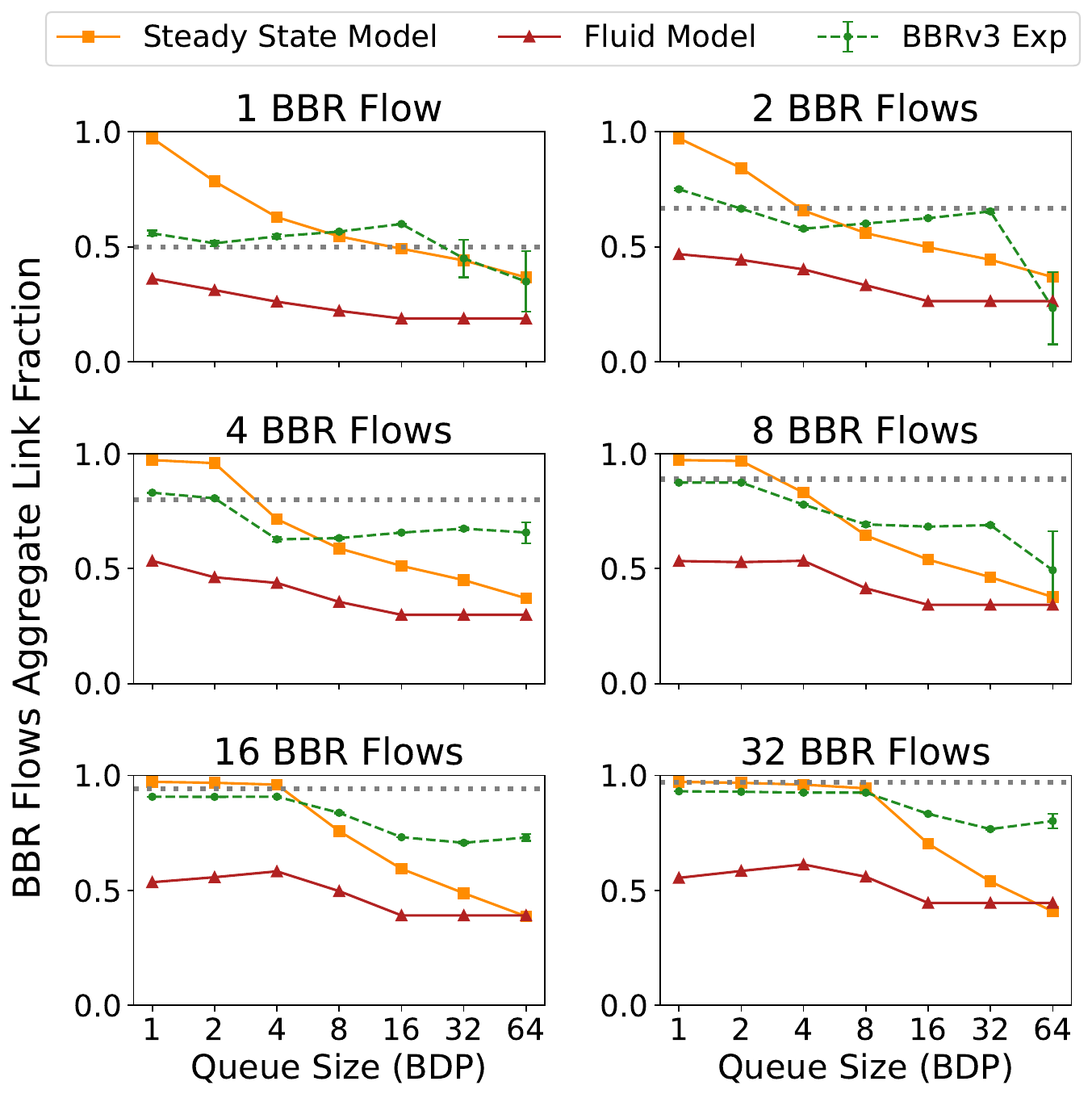}
      \caption{40ms × 10 Mbps, BBRv3 vs 1 Reno}
      \label{fig:part12-bbrv3-c}
  \end{subfigure}
  \caption{Aggregate link fraction captured by BBRv3 flows sharing a bottleneck with a loss-based flow. The gray dashed line indicates fair sharing. We compare results to BBRv1's steady-state model~\cite{2019IMC} and BBRv2's fluid model~\cite{scherrer}.} 
  \label{fig:part2-ware-settings-bbrv3}
\end{figure*}

\begin{figure*}[t]
    \centering
    \includegraphics[width=0.99\textwidth]{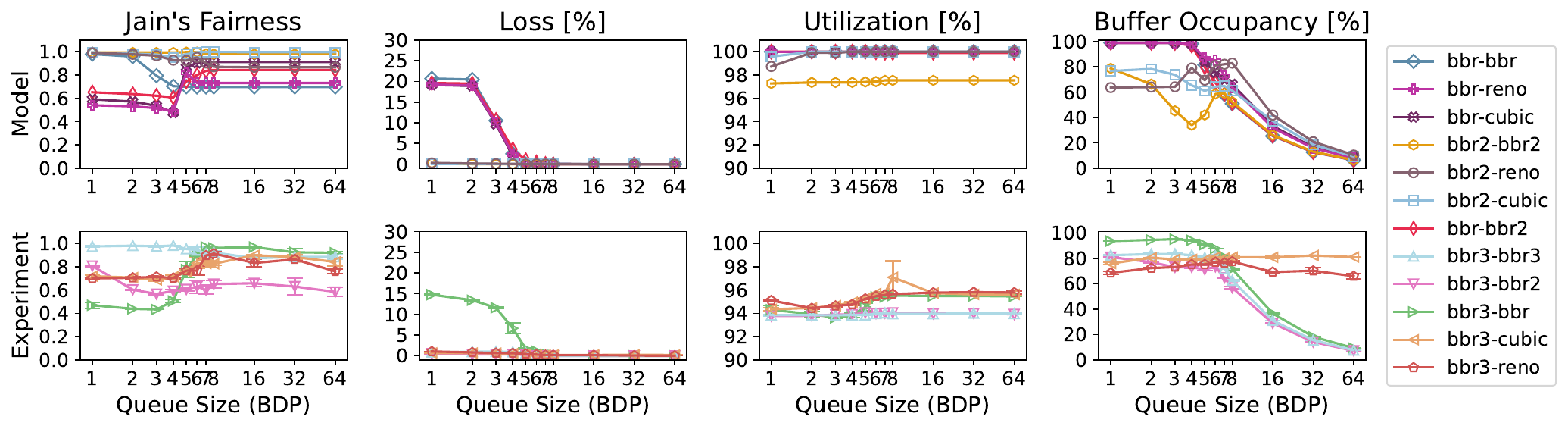}
    \caption{Fluid model prediction and BBRv3 experiment results when BBR shares a bottleneck with an equal number of flows of another type (different BBR version or loss-based CC), or when all flows sharing the bottleneck are the same.}
    \label{fig:part2-bbrv3-2022settings}
\end{figure*}

\subsection{Evaluating BBRv3 against single or multiple flows}

\textbf{BBRv3 against a single loss-based flow:} Fig.~\ref{fig:part2-ware-settings-bbrv3} shows the aggregate fraction of the link capacity obtained by BBRv3 flows when competing with a loss-based flow for various link settings and bottleneck buffer sizes. We also include the BBRv1's steady-state model~\cite{2019IMC} and the BBRv2's fluid model~\cite{scherrer} to evaluate their compatibility with BBRv3 experiment results. 

In shallow buffer scenarios, BBRv3 closely matches its fair share of link capacity, using more than BBRv2 but less than BBRv1. Unlike BBRv2, BBRv3 continues probing after a loss event, making it more aggressive. BBRv3 thus provides a balance between the aggressiveness of BBRv1 and the caution of BBRv2, as is evident in our results. BBRv3's throughput share also falls between the predictions of BBRv1's steady state model and BBRv2's fluid model.

In shallow buffers with few flows, the fluid model more closely matches BBRv3’s results than the steady-state model, while with many flows, the steady-state model is closer. In shallow buffers with many flows, the fluid model tends to under-predict BBRv3’s link share, while the steady-state model over-predicts with fewer flows. This reflects BBRv3’s balanced behavior between BBRv1 and BBRv2 — it behaves more like BBRv2 with fewer flows and small buffers but aligns with BBRv1 when there are many flows.

In deep buffer scenarios, BBRv3, like BBRv1 and BBRv2, captures less than its fair share of capacity, highlighting the dominance of CUBIC flows. However, when competing with less aggressive Reno flows, BBRv3 achieves a larger share, similar to our observations for BBRv2. Neither model closely predicts BBRv3’s throughput in CUBIC or Reno scenarios.

Recent experimental findings~\cite{pam2024} (under different network conditions) show that BBRv3 fairness is worse than that of BBRv1 when the buffer size is 1 BDP, but our results in Fig.~\ref{fig:part2-ware-settings-bbrv3} suggest the opposite. On the other hand, another study~\cite{bbrv3-mdpi} reports that BBRv3 shows better fairness than BBRv2 when competing with CUBIC. We believe this is due to RTT differences: in~\cite{pam2024}, the experiment base RTT is 100ms, which is higher than in our experiments. A 1BDP buffer size causes frequent losses, and after loss, CUBIC decreases its congestion window. At high RTTs, it takes a long time for CUBIC to ramp up, and in the mean time BBRv3 could dominate CUBIC. However, at lower RTTs (e.g., 30 ms), CUBIC recovers quickly and domination is not observed.
This motivates further experiments across a broader range of network settings to better characterize the protocol's behavior.

\begin{tcolorbox}[beforeafter skip=0.5\baselineskip, before upper={\parindent15pt},colback=black!3!white,colframe=black,
]

Neither model~\cite{2019IMC} nor~\cite{scherrer} is designed for BBRv3. However, the steady-state model of BBRv1~\cite{2019IMC} is a reasonable predictor of BBRv3 behavior when many flows compete in a shallow buffer, where both BBRv1 and BBRv3 tend to be aggressive against a loss-based flow. When fewer flows share a shallow buffer, our experiments suggest that BBRv3 strikes a balance between the aggressiveness levels of BBRv1 and BBRv2, so neither model is very predictive, and neither model describes BBRv3 behavior in deep buffers.

\end{tcolorbox}

\textbf{BBRv3 against multiple flows:} In this section, we evaluate the performance of BBRv3 in network settings shown in Fig.~\ref{fig: topology}b, and also analyze the compatibility of the fluid model with this version (refer to the second row in Figure~\ref{fig:part2-bbrv3-2022settings}).

\textbf{Fairness:} For intra-flow fairness, BBRv3 shows good fairness in both shallow and deep buffers—unlike BBRv1 or BBRv2. This deep-buffer behavior is consistent with findings in~\cite{pam2024,bbrv3-mdpi}. The fluid model for BBRv2 intra-flow fairness also shows high fairness across buffer sizes.

BBRv1 dominates BBRv3 in shallow buffer experiments, while their coexistence is better in deep buffers, similar to the case of BBRv1 vs. loss-based flows~\cite{pam2024}. The fluid model of BBRv1 vs. BBRv2 also reflects BBRv1's aggressive nature, which leads to unfairness in shallow buffers. 

BBRv3 vs. BBRv2 exhibits fairness issues, especially in deep buffers. In contrast, BBRv3 vs. BBRv1 and BBRv1 vs. BBRv2 show better fairness. This behavior has been documented in~\cite{bbrv3-wired} and confirmed by our experiments.

While competing against loss-based CCAs, BBRv3 shows slightly worse fairness than BBRv2 but better than BBRv1. However, Figure~\ref{fig:part2-ware-settings-bbrv3} indicates that BBRv3 is the most fair against loss-based flows. This discrepancy likely arises from differences in network conditions and evaluation methods. Specifically, Figure~\ref{fig:part2-bbrv3-2022settings} measures Jain’s Fairness Index with 5 BBRv3 and 5 Reno/CUBIC flows, while Figure~\ref{fig:part2-ware-settings-bbrv3} computes the BBRv3 share in a bottleneck with a single loss-based flow. Although theoretical models do not match this exact behavior, BBRv1 vs. loss-based predictions show a similar trend — less fairness in shallow buffers and better fairness in deep buffers.

\textbf{Loss:} In the BBRv3 vs. BBRv1 competition, shallow buffers experience high loss, confirming that BBRv1 consistently shows high loss in shallow buffers, whether against other BBR versions or loss-based CCAs. In contrast, other BBRv3 scenarios show minimal loss, similar to BBRv2, which also uses packet loss as a congestion signal.

\textbf{Utilization:} BBRv3 achieves higher utilization against loss-based CCAs than in BBRv3 vs. BBR scenarios, consistent with the trend in BBRv1 and BBRv2, where competition with loss-based CCAs improves bandwidth utilization.

\textbf{Buffer Occupancy:}
As the bottleneck buffer size increases, the buffer occupancy for BBR vs. BBR scenarios decreases, while it remains consistently high for BBR vs. loss-based CCA competition. The fluid model is not predictive for this as explained in Sec~\ref{BBR_multiple_flows}.

\begin{tcolorbox}[beforeafter skip=0.5\baselineskip, before upper={\parindent15pt},colback=black!3!white,colframe=black,
] 

BBRv3 shows better intra-flow fairness in shallow buffers than BBRv1, but it has inter-flow fairness issues with BBRv2. High losses persist with BBRv1 in shallow buffers. While fluid models for BBRv1 and BBRv2 partially explain BBRv3’s shallow buffer behavior, they fail in deep buffers—especially for intra-flow fairness, utilization, and buffer occupancy. Existing models do not fully capture BBRv3’s performance, indicating the need for a new model.

\end{tcolorbox}

\section{Conclusion}\label{conclusion}

In this work, we evaluate the theoretical models from two influential papers about TCP BBR's behavior over a shared bottleneck, and compare with our experiment results.

Our work confirms the utility of the steady state model~\cite{2019IMC} and the fluid model~\cite{scherrer} for BBR in specific scenarios despite the updates to both BBRv1 and BBRv2 since these papers were published. We find that the steady state model is more accurate than the fluid model at predicting the aggregate link fraction captured by many BBR flows when sharing the link with a loss-based flow. The fluid model demonstrates good accuracy in capturing key dynamics of BBRv1 and BBRv2 for shallow to moderate buffer sizes, but shows some limitations when applied to scenarios with deep buffers, a large number of flows, or complex intra-flow fairness interactions. This is particularly relevant since home routers can have deep buffers, as noted in~\cite{ware_reference_high_buffer}.

We similarly confirm that for buffer sizes up to 7~BDP, the fluid model~\cite{scherrer} for BBR explains the experimental behavior of BBRv1 and BBRv2 when sharing a bottleneck link with other flows of similar or dissimilar type, with a few exceptions. 

Although both models predate BBRv3, since BBRv3 is a minor update to BBRv2, we might expect that some of the findings may still apply. However, our experiments show that neither model fully captures the behavior of the recently introduced BBRv3, particularly in deep buffer settings and scenarios involving numerous flows. While some aspects of BBRv3 align closely with predictions from the fluid model, significant discrepancies remain in intra-flow fairness, buffer occupancy, and link utilization under certain conditions. These discrepancies indicate that existing theoretical models may require updates—or entirely new modeling approaches—to accurately represent BBRv3’s behavior.

Our experiment artifacts are publicly available~\cite{our_artifacts_github_repo} so that others may validate and build on these findings. 

While our results confirm some important findings, they also disagree in some instances, and the behavior of BBRv3, in particular, needs to be better understood. Importantly, our study is limited to bulk-transfer traffic patterns and specific network settings. As a result, it remains unclear how well these findings generalize to more realistic application behaviors, or to a broader range of network scenarios, including higher link capacities, and diverse RTTs. These results and limitations suggest the need for additional, more comprehensive experiments and modeling, which we plan to undertake in future work.

\section*{Acknowledgments}
This research was supported by the New York State Center for Advanced Technology in Telecommunications (CATT), NYU WIRELESS, and the National Science Foundation (NSF) under Grant No. CNS-2148309 and OAC-2226408.
\vspace{-0.05cm}

\bibliographystyle{IEEEtran}
\bibliography{reference}

\end{document}